\documentclass{article}
\usepackage{graphicx,amsmath,amssymb,bm}

\def\abstract#1{\vskip 7mm 
        \begin{center}{\large Abstract}\par \smallskip
                \begin{minipage}[c]{12cm}
                        \small #1
                \end{minipage}
        \end{center}
}
\def\title#1{\begin{center}{\Large\bf #1}\end{center}}
\def\author#1{\vskip 5mm \begin{center}{#1}\end{center}}
\def\address#1{\begin{center}{\it #1}\end{center}}
\makeatletter
\@ifundefined{lesssim}{}{}
\@ifundefined{gtrsim}{}{}
\def\vereq#1#2{\lower3pt\vbox{\baselineskip1.5pt \lineskip1.5pt
\ialign{$\m@th#1\hfill##\hfil$\crcr#2\crcr\sim\crcr}}}
\makeatother

\newcommand{\nn}{\nonumber\\}

\newlength{\minitwocolumn}
\begin{document}

\title{%
An Investigation of the Tomimatsu-Sato Spacetime
}
\author{%
  Wataru Hikida\footnote{E-mail:hikida@yukawa.kyoto-u.ac.jp} and
  Hideo Kodama\footnote{E-mail:kodama@yukawa.kyoto-u.ac.jp}
}
\address{%
Yukawa Institute for Thoretical Physics, Kyoto University, Kyoto 606-8502,Japan
}
 
\abstract{
We investigate the structure of the $\delta=2$ Tomimatsu-Sato spacetime.
We show that this spacetime has degenerate horizons with two components, 
in contrast to the general belief that the Tomimatsu-Sato
solutions with even $\delta$ do not have horizons.
}

\section{Introduction}
One of the most important problems in the relativistic astrophysics and
gravitation theory is the final state of a massive star.
In general relativity, such a collapse is usually assumed to end up with
the formation of a regular black hole, but the possibility of the
formation of a naked singularity is not ruled out.
Concerning to this problem, Penrose \cite{Pe69} proposed the conjecture
that the latter possibility will not be realized under physically
realistic conditions, which is called the {\it cosmic censorship
hypothesis}. 
This conjecture, however, still remains to be proved and there are many
counterexamples in the spherically symmetrical cases.

The best known example of naked singularities is the shell-focusing
singularities in the spherically symmetric dust collapse, so called the
Tolman-Bondi model \cite{To34,Bo47}.
Another important example is the critical phenomena in gravitational
collapse, which was first found for spherically symmetric collapse of a
massless scalar field by Choptuik \cite{Ch93}.
In this example, although the naked singularity formation is not
generic, a naked singularity solution and a self-similar structure
appears as a critical point for a generic continuous initial data.
Hence, there exist a generic set of regular solutions in any small
neighborhood of the singular solutions in the initial data space.
Due to this, the structure of the singular solutions play a crucial role
in determining the behavior of solutions near the critical point, and the
fate of the gravitational collapse.

Although these examples show that the formation of naked singularities
is rather generic and solutions with naked singularities may play a key
rule for the spherically symmetric gravitational collapse, the assumption
of the spherically symmetry is too restrictive to say anything on the
cosmic censorship hypothesis in realistic systems.
We must investigate systems with less symmetry.
In such investigations, it is expected that the analysis of naked singular
solutions provides useful information on the final fate of gravitational
collapse, as the above example of the critical phenomena indicates.
From this point of view, the Tomimatsu-Sato solutions are the best system
to analyze as the starting point, because they are the simplest
extension of the Kerr black hole solution.

Although it has been a long time since the Tomimatsu-Sato solutions were
found, little is known about these solutions yet and sometimes wrong
statements were made in the literature.
For example, although it has been shown that the Tomimatsu-Sato solutions
with odd $\delta$ have a Killing horizon at the segment $\rho=0$,
$|z|<\sigma$ in the Weyl coordinate, it has been believed that the
solutions with even $\delta$ have no horizon because the corresponding
segment for these solutions is not Killing horizon \cite{GiCl73}.
In the present paper, we will show that this belief is false and the two
points $\rho=0,z=\pm \sigma$ are actually degenerate horizons for
$\delta=2$.

\section{Tomimatsu-Sato Spacetime}
In canonical form of the stationary axisymmetric metric \cite{Kra80}
\begin{equation}
 ds^2=-f(dt-\omega d\phi)^2+f^{-1}[e^{2\gamma}(d\rho^2+dz^2)+\rho^2d\phi^2],
\end{equation}
the Kerr-TS family is expressed as \cite{TS73}
\begin{equation}
f=\frac{A}{B} ,\quad \omega=\frac{2mq}{A}(1-y^2)C ,
\quad e^{2\gamma}=\frac{A}{p^{2\delta}(x^2-y^2)^{\delta^2}},
\end{equation}
where the functions $A,B$ and $C$ are polynomials with the degrees of
$2\delta^2,2\delta^2$ and $2\delta^2-1$, respectively, in the prolate
spheroidal coordinates $x$ and $y$ defined by
\begin{equation}
  \rho=\sigma\sqrt{(x^2-1)(1-y^2)},\quad z=\sigma xy,
\end{equation}
and $\sigma,p$ and $q$ are related with mass $m$ and angular momentum
$J$ as
\begin{equation}
 p^2+q^2=1,\quad \sigma=\frac{mp}{\delta},\quad J=m^2q.\label{Feb 13 22:05:03 2003}
\end{equation}
In the Kerr ($\delta=1$) and the $\delta=2$ Tomimatsu-Sato case, $A,B$
and $C$ are given by
\begin{eqnarray}
 \delta=1 &:& A=p^2(x^2-1)-q^2(1-y^2),\quad B=(px+1)^2+q^2y^2,
\quad C=-px-1,\\
 \delta=2 &:&
A=p^4(x^2-1)^4+q^4(1-y^2)^4\nonumber\\
  &&\qquad  -2p^2q^2(x^2-1)(1-y^2)\{2(x^2-1)^2+2(1-y^2)^2+3(x^2-1)(1-y^2)\}\nonumber\\
 &&B=\{p^2(x^2+1)(x^2-1)-q^2(y^2+1)(1-y^2)+2px(x^2-1)\}^2\nn
  &&\qquad  +4q^2y^2\{px(x^2-1)+(px+1)(1-y^2)\}^2\nonumber\\
 &&C=-p^3x(x^2-1)\{2(x^2+1)(x^2-1)+(x^2+3)(1-y^2)\}\nn
  &&\qquad  -p^2(x^2-1)\{4x^2(x^2-1)+(3x^2+1)(1-y^2)\}+q^2(px+1)(1-y^2)^3.
\end{eqnarray}
In this paper, we focus on the $\delta=2$ Tomimatsu-Sato metric.
The properties of the corresponding spacetime that have been found so
far are summarized as follows \cite{TS73,GiCl73}.
\begin{description}
 \item[Ring singularity:] This spacetime has a ring singularity at the
	    root of $B(x,y=0)=0$. (The cross in Figure \ref{fig:property})
 \item[Ergosphere:] The timelike Killing vector becomes null at the roots
	    of $A$. There are two single roots for $x\geq1$ (The dotted lines in
	    Figure \ref{fig:property}), the smaller of
	    which coincides with the ring singularity. 
 \item[Causality violation region:] Since $\phi$ is the periodic angular
	    coordinate, this spacetime has closed timelike
	    loops in the region with negative $g_{\phi\phi}$. (The shaded portion
	    in Figure \ref{fig:property})
 \item[Directional singularity:] This metric has directional
	    singularities at the points $(\rho,z)=(0,\pm\sigma)$,
	    where the value of a curvature
	    invariant has different limits when the point is approached from
	    different directions. (The two filled circles in Figure
	    \ref{fig:property})
 \item[The nature of the surface $x=1$:] In the Kerr case,
	    $x=1$($\rho=0,|z|\leq\sigma$) surface is an event
	    horizon. However, in the $\delta=2$ Tomimatsu-Sato case, this
	    is not
	    the case. Because the induced metric is Lorentzian and two
	    Killing vectors $\partial_t$ and $\partial_\phi$ become
	    parallel there, $x=1$ surface cannot be a null surface.
\end{description}

\setlength{\minitwocolumn}{0.5\textwidth}
\addtolength{\minitwocolumn}{0.3 \columnsep}
\begin{figure}[t]
  \begin{tabular}{c c}
    \begin{minipage}{\minitwocolumn}
      \begin{center}
 \resizebox{!}{6cm}{\includegraphics{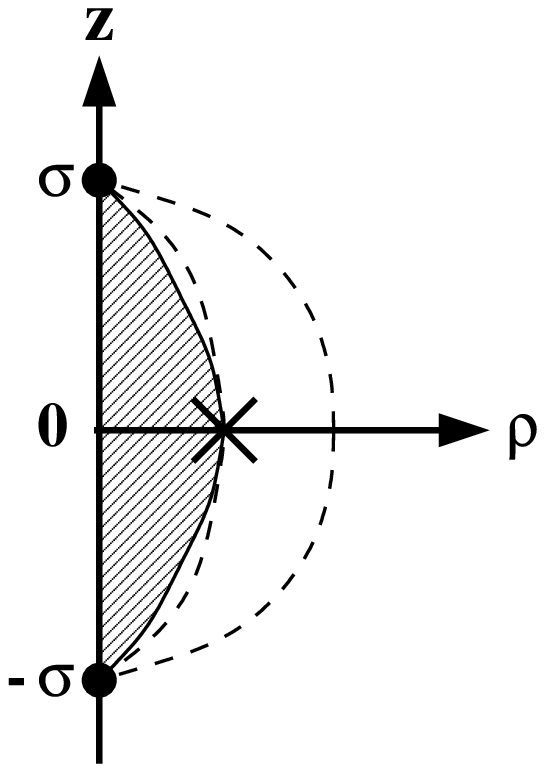}}
	 \caption{The structure of the spacetime. The crossed sign, dotted
       lines and the two filled circles denote the ring singularity, the infinite redshift
       surfaces and the directional
	    singularities respectively. 
In the shaded portion, the closed timelike
       loops exist.}
	 \label{fig:property}
      \end{center}
    \end{minipage}
    &
    \begin{minipage}{\minitwocolumn}
      \begin{center}
 \resizebox{!}{6cm}{\includegraphics{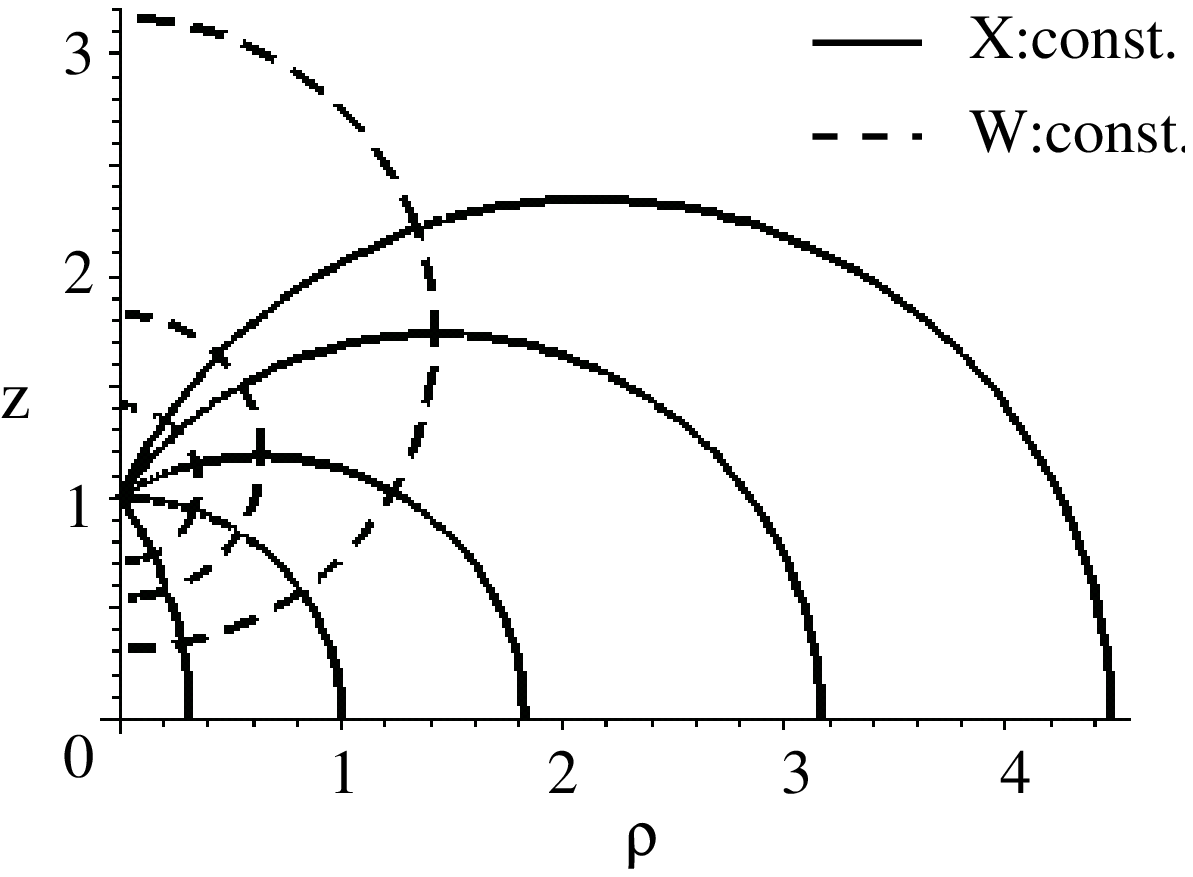}}
	 \caption{The relation between $(\rho,z)$ and $(X,W)$
       coordinate system. The solid lines and the dotted lines are
       $X=$constant and $W=$constant lines respectively.}
	 \label{fig:property2}
      \end{center}
    \end{minipage}
  \end{tabular}
\end{figure}

\section{Extension of Tomimatsu-Sato spacetime}
Ernst \cite{Ern76} pointed out that the quasi-regular directional
singularities at $\rho=0,z=\pm\sigma$ are hypersurfaces in reality, by
introducing a polar coordinate system around each point.
He further studied the behavior of geodesics crossing these surfaces,
and concluded that they are timelike surfaces.
Now, we show that his conclusion is false and these surfaces are horizons,
by introducing an appropriate coordinate system.

First, we rewrite the metric in the former
\begin{equation}
 ds^2=-e^{2\nu}dt^2+e^{2\psi}(d\phi-\Omega dt)^2
      +e^{2\mu}\left(\frac{dx^2}{x^2-1}+\frac{dy^2}{1-y^2}\right),
\end{equation}
where
 \begin{eqnarray}
 e^{2\nu}&=&\frac{p^2(x^2-1)B}{D}\ ,\ \Omega=-\frac{4pqC}{D}\ 
,\ e^{2\psi}=\frac{(1-y^2)D}{p^2B},\ 
 e^{2\mu}=\frac{B}{p^{4}(x^2-y^2)^{3}}\label{Dec 30 22:19:15 2002},
\end{eqnarray}
and $D$ is a polynomial determined by the equation
       \begin{equation}
	D=\frac{p^2(x^2-1)B^2-4\delta^2q^2(1-y^2)C^2}{A}.
       \end{equation}
In the $\delta=2$ Tomimatsu-Sato case, the explicit expression for $D$
is given by
\begin{eqnarray}
	D&=&p^6(x^2-1)(x^8+28x^6+70x^4+28x^2+1)-16q^6(1-y^2)^3\nn
	&+&p^4q^2[(x^2-1)\{32x^2(x^4+4x^2+1)-4(1-y^2)(x^2-1)^3+(-6x^4+12x^2+10)(1-y^2)^3\}\nn
	&-&4(1-y^2)^3(x^4+6x^2+1)]+p^2q^4[(x^2-1)\{64x^4+(1-y^2)^2(y^4+14y^2+1)\}\nonumber\label{Dec 30 22:41:36 2002}\\
	&-&16(1-y^2)^3(x^2+2)]+8p^5x(x^4-1)(x^4+6x^2+1)-32pq^4x(1-y^2)^3\nonumber\\ 
	&+&8p^3q^2x[(x^2-1)\{8x^2(x^2+1)+(1-y^2)^2(2y^2-x^2+1)\}-4(1-y^2)^3].
       \end{eqnarray}
Here, we have renormalized $\sigma$ as $\sigma=1$.

Next, we introduce the new coordinates $X$ and $W$ by
\begin{eqnarray}
 x^2 = \dfrac{1+X}{1+X-W},\qquad
 y^2 = \dfrac{(1+X)(1-W)}{1+X-W}
\end{eqnarray}
Figure \ref{fig:property2} shows the relation between the $(\rho,z)$ and
the $(X,W)$ coordinate system.
In this $(X,W)$ coordinate system, the metric is written
\begin{equation}
  ds^2=-e^{2\nu}dt^2+e^{2\psi}(d\phi-\Omega dt)^2+e^{2\mu}\frac{(1+X)W}{4(1+X-W)}
\left[\frac{dX^2}{(1+X)^2X}+\frac{dW^2}{(1-W)W^2}\right].
\end{equation}
In the region near $W=0$, $A,B,C$ and $D$ behave as
\begin{eqnarray}
 A
  &=&\left(\frac{W}{1+X-W}\right)^4
\left(p^4+X^4q^4-4p^2q^2X-4p^2q^2X^3-6p^2q^2X^2\right),\nonumber\\
B
 &\sim& 8(p+1)(p^2+q^2X^2)\biggl(\frac{W}{1+X-W}\biggr)^2 
       +8(p+1)(p^2-q^2X^3)\biggl(\frac{W}{1+X-W}\biggr)^3,\nonumber\\
C
 &\sim&-4p^2(1+X)(p+1)\left(\frac{W}{1+X-W}\right)^2,\nn
 D
&\sim& 64p^2(p+1)^2\left(\frac{W}{1+X-W}\right)
     +128p^2(p+1)^2\left(\frac{W}{1+X-W}\right)^2.
\end{eqnarray}
From this, it follows that at $W\sim0$, $e^{2\nu},e^{\psi},\Omega$ and
$e^\mu$ are expressed as
\begin{eqnarray}
  e^{2\nu}
           &\sim& \frac{p^2+q^2X^2}{8(p+1)}\biggl(\frac{W}{1+X}\biggr)^2
         +\frac{-q^2X^3+p^2}{8(p+1)}\left(\frac{W}{1+X}\right)^3,\nonumber\\
\Omega &\sim& \frac{pq(1+X)}{4(p+1)}\left(\frac{W}{1+X-W}\right),\nn
  e^{2\psi}&\sim& \frac{8(p+1)X}{(p^2+q^2X^2)}, \nonumber\\
  e^{2\mu} &\sim& \frac{8(p+1)(p^2+q^2X^2)}{p^4W(1+X)^2}
       +\frac{8(p+1)(-q^2X^2-q^2X^3)}{p^4(1+X)^3}.
\end{eqnarray}
This asymptotic behavior suggests that the $\delta=2$ Tomimatsu-Sato
spacetime can be extended regularly across the surface $W=0$ in terms
of the advanced coordinate $v$ and the $\phi_+$ coordinate defined by
\begin{equation}
dv=dt+\frac{4(p+1)dW}{p^2W^2\sqrt{1-W}},\quad  
d\phi_+=d\phi+\frac{qdW}{pW\sqrt{1-W}}.
%
%
\end{equation}
Actually, in the $(v,W,X,\phi_+)$ coordinate system, the metric is
expressed as
\begin{eqnarray}
 ds^2&=&-e^{2\nu}\left(dv-\frac{4(p+1)dW}{p^2W^2\sqrt{1-W}} \right)^2
        +e^{2\psi}\left\{\left(d\phi_+ -\frac{qdW}{pW\sqrt{1-W}} \right) 
        - \Omega \left(dv-\frac{4(p+1)dW}{p^2W^2\sqrt{1-W}}\right)\right\}^2\nonumber\\
       &&+e^{2\mu}\frac{(1+X)W}{4(1+X-W)}
         \left[\frac{dX^2}{(1+X)^2X}+\frac{dW^2}{(1-W)W^2}\right]\nonumber\\
 &=& -\frac{p^2+q^2X^2}{8(p+1)}\left(\frac{W}{1+X}\right)^2[1+O(W)]dv^2
+\frac{p^2+q^2X^2}{p^2(p+1)}\frac{1+O(W)}{(1+X)^2}dvdW\nonumber\\
&&+\frac{8(p+1)X}{(p^2+q^2X^2)}[1+O(W)]
\left(d\phi_+ -\frac{pqW}{4(p+1)}dv+O(W^2)dv+O(1)dW\right)^2\nonumber\\
&&+O(1)dW^2+\frac{2(p+1)(p^2+q^2X^2)}{p^4(1+X)^2}[1+O(W)]\frac{dX^2}{(1+X)^2X},
\end{eqnarray}
which is regular at $W=0$.
This expression also shows that $W=0$ surface is a null hypersurface,
whose normal vector is a Killing vector.
Hence, the $W=0$ surface is a Killing horizon.
Because $W=0$ is the double root of $e^{2\nu}$, the horizon is
degenerate, i.e., the surface gravity vanishes.

Finally, we examine the shape of horizon. 
The Tomimatsu-Sato spacetime has a Killing vector $\partial_\phi$.
So the radius $R$ is given by
\begin{equation}
 g_{\phi\phi}|_{(X_0,W_0)}=R^2|_{(X_0,W_0)}.
\end{equation}
Figure \ref{fig:graph} shows the relation between $g_{\phi\phi}$ and $X$
at horizon ($W=0$).
There $p^2$ parameterizes the angular momentum as is seen from
eq.(\ref{Feb 13 22:05:03 2003}): $p^2=1$ corresponds to the non-rotating
case, and $p^2$ decreases as the angular momentum increases.
Figure \ref{fig:tshorizon} expresses the shape of horizon for $p^2=0.5$.
From this figure we see that the horizons of $\delta=2$ Tomimatsu-Sato
spacetime consist of two spheres.

\setlength{\minitwocolumn}{0.5\textwidth}
\addtolength{\minitwocolumn}{0.3 \columnsep}
\begin{figure}[t]
  \begin{tabular}{c c}
    \begin{minipage}{\minitwocolumn}
      \begin{center}
 \resizebox{!}{5cm}{\includegraphics{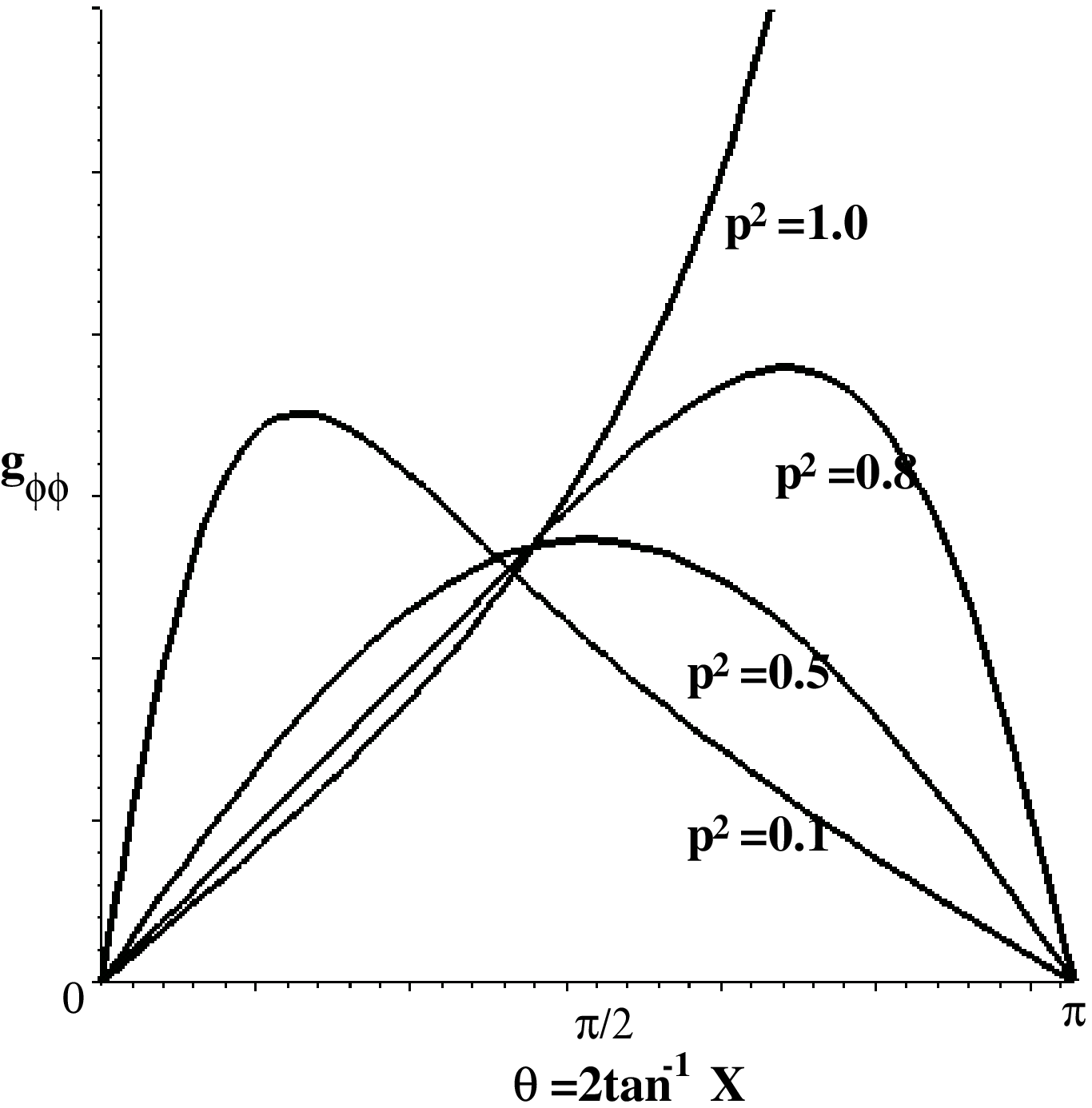}}
	 \caption{The relation between $g_{\phi\phi}$ and
$X$ at horizon ($W=0$).}
	 \label{fig:graph}
      \end{center}
    \end{minipage}
    &
    \begin{minipage}{\minitwocolumn}
      \begin{center}
 \resizebox{!}{5cm}{\includegraphics{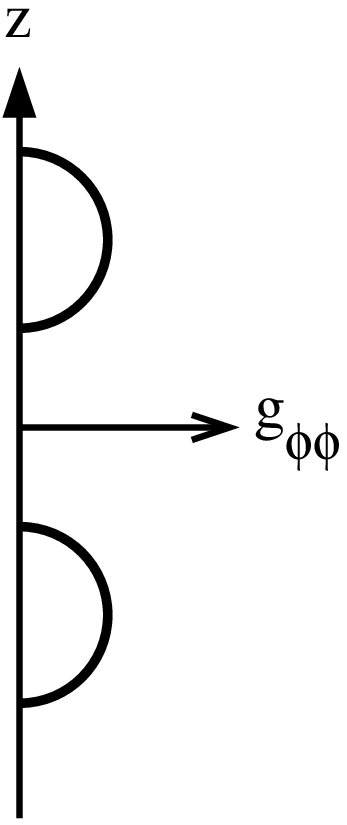}}
	 \caption{The shape of horizon in the $p^2=0.5$ case.}
	 \label{fig:tshorizon}
      \end{center}
    \end{minipage}
  \end{tabular}
\end{figure}

\section{Conclusion}
We have investigated the structure of the $\delta=2$ Tomimatsu-Sato spacetime.
By introducing an appropriate coordinate system, we have shown that the
two points in the Weyl coordinates, which have been recognized as
the directional singularities, are really two-dimensional surfaces and that
these surfaces are horizons.
We have also shown that each of the two horizons has the topology of a
sphere.
This result is rather surprising because the $\delta=2$ Tomimatsu-Sato
solution is obtained from the Neugebauer-Kramer solution representing a
superposition of two Kerr solutions, as the limit that the centers of two
black holes coincide \cite{OS81}.
This may indicates a new possibility for the final states of
gravitational collapse.


\begin{thebibliography}{99}
\bibitem{Pe69} R.Penrose Riv.\ Nuovo.\ Cimento.\  {\bfseries 1}, 252
	 (1969)
 \bibitem{To34} R.C.Tolman Proc.\ Nat.\ Acad.\ Sci.\ {\bfseries 20}, 169
	 (1934)
 \bibitem{Bo47} H.Bondi Mon.\ Not.\ R.\ Astron.\ Soc.\ {\bfseries 107},
	 410 (1947)
 \bibitem{Ch93} M.W.Choptuik Phys.\ Rev.\ Lett.\ {\bfseries 70}, 9
	 (1993)
\bibitem{Kra80} D.~Kramer, H.~Stephani, M.~MacCallum and E.~Herlt eds.:
	{\it Exact Solutions of Einstein's Field Equations}(Cambridge
	Univ. Press, Cambridge, 1980)  
 \bibitem{TS73} Akira Tomimatsu and Humitaka Sato 
Prog.\ Theor.\ Phys.\  {\bf 50}, 95 (1973)
 \bibitem{GiCl73} G.W.Gibbons and R.A.Russell-Clark. Phys.\ Rev.\ Lett.\
	{\bf 30}, 398 (1973) 
 \bibitem{Ern76} F.J.Ernst. J.\ Math.\ Phys.\ {\bf 17}, 1091 (1976)
\bibitem{OS81} Ken-ichi Oohara and Humitaka Sato. Prog.\ Theor.\
	 Phys.\ {\bf 65}, 1891 (1981)
\end{thebibliography}
\end{document}